# scientific reports

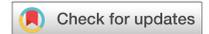

OPEN

# Bimagnon dispersion of $La_2CuO_4$ probed by resonant inelastic X-ray scattering

A. Singh[1,2], H. Y. Huang[1], K. Tsutsui[3], T. Tohyama[4], S. Komiya[5], J. Okamoto[1], C. T. Chen[1], A. Fujimori[1,6,7] & D. J. Huang[1,8,9]✉

We report on the study of the magnetic excitations of Mott insulator $La_2CuO_4$ by using resonant inelastic x-ray scattering (RIXS) and cluster calculations within the framework of exact diagonalization. Our results demonstrate experimentally that the bimagnon excitation in Cu $L_3$-edge RIXS is enhanced if the incident x-ray energy is slightly above the absorption edge. Through incident-energy-dependent momentum-resolved RIXS, we investigated the excitation of the bimagnon with predominantly $A_{1g}$ symmetry. The bimagnons of $La_2CuO_4$ exhibit a nearly flat dispersion with momentum along the Cu-O bond direction. This observation agrees with the bimagnon dispersion from the calculations on a single-band Hubbard model rather than a Heisenberg model with only the nearest neighbor exchange interaction. This means that the effect of the higher-order spin couplings such as the cyclic or ring exchange interactions caused by the coherent motion of electrons beyond nearest-neighbor sites is important for understanding the bimagnon dynamics of cuprates.

While the mystery of novel cuprate superconductivity is unresolved, spin fluctuations are still considered to play an important role in its pairing mechanism. Spin fluctuations around the antiferromagnetic wave vector could contribute to the pairing interaction[1–6]. The interplay between spin and charge degrees of freedom could be the key to resolving the mystery[6,7]. Understanding the two-dimensional magnetic properties of its mother compound is crucial to describing the metallic behavior of a doped cuprate[8–10]. The Heisenberg model with only the nearest-neighbor exchange interaction $J$ is a conventional starting point for the two-dimensional spin-$\frac{1}{2}$ antiferromagnet[11]. This simple model, however, gives reduced spectral weight of magnons along the antiferromagnetic zone boundary[12,13], and corrections with long-range exchange interactions[12,14] or higher-order exchange interactions such as ring exchange[12] are needed. In contrast, the Hubbard model, which comprises the on-site Coulomb energy $U$ and the hopping energy $t$ between nearest-neighbor Cu sites, is expected to naturally include the effect of such higher-order interactions through the coherent motion of electrons beyond nearest-neighbor sites[15–17].

There have been extensive studies on the spin fluctuations of undoped cuprates such as $La_2CuO_4$[18], particularly the measurements with inelastic neutron scattering (INS)[12,13]. In past decades, much experimental evidence showed that resonant inelastic x-ray scattering (RIXS) is a complementary tool to INS for probing magnetic fluctuations[19–25]. Theoretically, direct spin-flip scattering is allowed in $L$-edge RIXS[26–32]; Jia et al. showed that Cu $L$-edge RIXS provides access to the spin dynamical structure factor if the x-ray polarization is fully analyzed[30,31]. The pioneering work of Braicovich et al. on Cu $L$-edge RIXS has successfully measured the single-magnon dispersion of cuprates[19–21]. With crossed polarizations between the incident and scattered x-rays, RIXS detects spin fluctuations directly, requiring a much smaller sample volume and enabling easy detection of high-energy excitations, as opposed to INS. For some cases, it is even more powerful than INS as it also probes charge, orbital, and lattice excitations and the effect of the matrix element gives rise to a possibility of tuning the cross-section involved with different scattering channels; for example, see Refs. [22,23,33–41]. However, separating bimagnon or even-order excitation from a single magnon is delicate in Cu $L_3$-edge RIXS[42]. Using exact diagonalization (ED) calculations on a single-band Hubbard model, Tsutsui and Tohyama[32] revealed that

[1]National Synchrotron Radiation Research Center, Hsinchu 30076, Taiwan. [2]Department of Physics and Astrophysics, University of Delhi, New Delhi 110007, India. [3]Synchrotron Radiation Research Center, National Institutes for Quantum Science and Technology, Hyogo 679-5148, Japan. [4]Department of Applied Physics, Tokyo University of Science, Tokyo 125-8585, Japan. [5]Central Research Institute of Electric Power Industry, Yokosuka, Kanagawa 240-0196, Japan. [6]Center for Quantum Science and Technology and Department of Physics, National Tsing Hua University, Hsinchu 30013, Taiwan. [7]Department of Physics, University of Tokyo, Bunkyo-ku, Tokyo 113-0033, Japan. [8]Department of Physics, National Tsing Hua University, Hsinchu 30013, Taiwan. [9]Department of Electrophysics, National Yang Ming Chiao Tung University, Hsinchu 30093, Taiwan. ✉email: djhuang@nsrrc.org.tw

Cu $L$-edge RIXS can measure bimagnon excitations of cuprates at half filling if the incident photon energy is tuned to be above the $L_3$ absorption energy in the order of $J$.

In the present article, we combined RIXS measurements and ED calculations on the single-band Hubbard model to study the magnetic excitations of $La_2CuO_4$, which represents a paradigm for quantum magnetism and is a parent compound of cuprate superconductors. For the first time, we demonstrate experimentally that bimagnon excitation in RIXS is enhanced if the incident x-ray energy is slightly above the absorption edge. The measured bimagnon dispersion of the in-plane momentum transfer $q_\parallel$ along the Cu-O bond direction agrees well with the ED calculations. This paper is organized as follows. We begin by describing the experimental and computational methods, including Cu $L$-edge RIXS measurements and ED calculations. Next, we present the Cu $L$-edge RIXS results, focusing on their dependence on incident energy and momentum. Finally, we discuss the observed bimagnon dispersions.

## Results

### Incident-energy-dependent RIXS

Figure 1 shows incident-energy-dependent RIXS spectra at Cu $L_3$-edge with $q_\parallel = (\pi, 0)$ in units of $1/a$, with which all momentum transfers are expressed throughout the paper; here, $a$ is the lattice constant. In the RIXS process of cuprates, a $2p$ electron is excited by an incident x-ray of energy $\omega_i$ to the unoccupied upper Hubbard band, followed by the decay of an electron in the valence $3d$ band to the $2p$ core hole to form a $dd$ excitation. Figure 1b plots a series of RIXS spectra from $La_2CuO_4$, when the incident energy $\omega_i$ is tuned to the $L_3$ absorption edge, the intensity of the $dd$ feature reaches a maximum. The energy and spectral line shape of the $dd$ excitation reflect the local electronic structure. We observed $dd$-excitation features at energies ∼ 1.5 eV, 1.7 eV, and 2.0 eV.

In addition, collective excitations such as phonon, magnon and plasmon[21,36,37,43,44] excitations can be induced by Cu $L$-edge RIXS, accompanying the $dd$ excitations. In the RIXS intermediate state, the $2p$ core hole is with strong spin-orbit coupling. As the spin is no longer a good quantum number in the $2p$ orbital, $L$-edge RIXS permits excitation involved with a spin flip. The matrix elements of the operators $S_q^j$ and $N_q^j$ govern the RIXS excitations with the change of total spin by one ($\Delta S = 1$) and no spin change ($\Delta S = 0$), respectively. If the polarization of the incident x-ray is selected, one can in principle analyze the polarization of the scattered x-ray to separate RIXS excitations with $\Delta S = 1$ and 0.

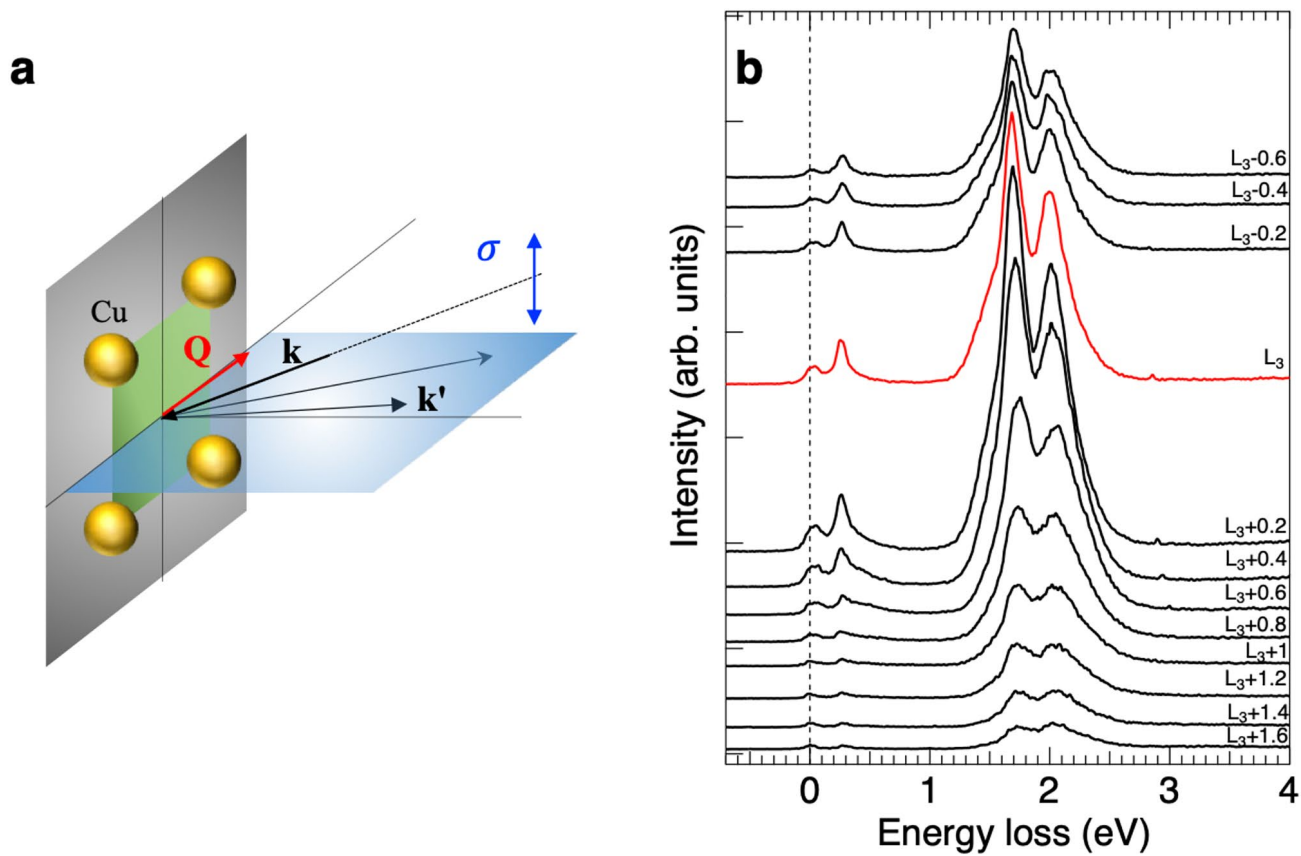

**Fig. 1.** Incident-energy-dependent Cu $L$-edge RIXS of $La_2CuO_4$. (**a**) Scattering geometry of the RIXS measurements. The scattering plane is defined by the incident and scattered wave vectors, **k** and **k′**, respectively. The projection of the momentum transfer onto the $CuO_2$ plane, $q_\parallel \equiv$ **k** − **k′** is along the antinodal direction $(\pi, 0)$. (**b**) RIXS spectra of $q_\parallel = (0.5\pi, 0)$ for various incident energies across Cu $L_3$-edge. The red spectrum corresponds to the $L_3$ peak in XAS. Spectra are plotted with a vertical offset for clarity.

Before presenting our RIXS data, we first discuss the resonance effect of the magnon and bimagnon excitations in RIXS through ED calculations. For the half-filling, a bimagnon-type operator dominates the non-spin-flip process[32]. Figure 2a, b compare the calculated RIXS excitations with $\Delta S = 1$ and 0 at selected in-plane momentum transfers for the incident photon energy $\omega_i$ tuned to the $L_3$ absorption edge and 0.27 eV above, defined as $\omega_0$. The comparison indicates that, although the intensity is weak, the bimagnon RIXS excitation has higher energy than that of a single magnon, roughly by 0.2 eV. Figure 2c plots calculated intensities of absorption spectrum (XAS) and RIXS with $\Delta S = 1$ and 0 as a function of the incident x-ray energy. The bimagnon RIXS resonates at an incident photon energy 0.27 eV higher than the XAS edge and the single-magnon resonance. To conceptually explain the reason that the bimagnon excitation is enhanced at energy slightly above the absorption edge, we depict the spin configurations around the $3d^{10}$ site of the XAS final state, i.e., the intermediate state of RIXS, in Fig. 2d, e for the incident energy tuned to the absorption edge and $\omega_0$, respectively.

In the initial state of the XAS and RIXS, there are configurations in which an up spin and a down spin are exchanged from the complete Néel spin configuration because of the quantum fluctuations. Figure 2d, e illustrate the spin configurations of the RIXS intermediate state (equivalent to the XAS final state) in two distinct scenarios. Figure 2d shows the excitation of an up-spin electron from the $2p$ core level to a down-spin site on the down-spin sublattice in the Néel configuration. In contrast, Fig. 2e depicts an alternative channel, in which an up-spin electron is excited to an exchanged down-spin site on the up-spin sublattice in the fluctuated configuration. The spin arrangements surrounding the core site differ between these two cases, resulting in an energy difference on the order of $J$. This energy difference can naturally lead to bimagnon excitations once the up-spin electron and the core hole are removed via photon emission. That is, X-rays with additional energy on the order of $J$ can enhance the bimagnon spectral weight. Therefore, the bimagnon excitation can enhance through a state in energy higher than the XAS edge state.

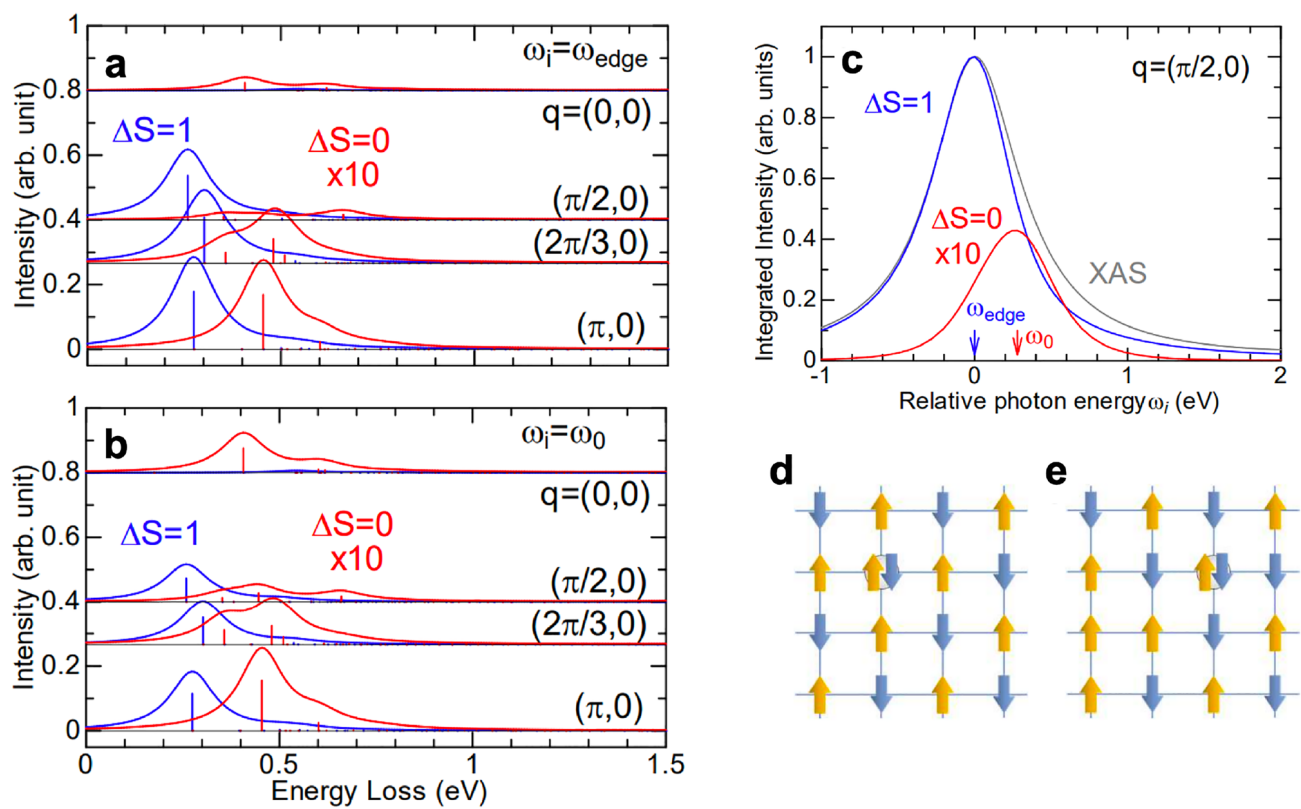

**Fig. 2.** Calculated Cu $L_3$-edge RIXS spectra with the spin-flip and non-spin-flip processes. (**a,b**) RIXS spectra for various in-plane momenta $\mathbf{q}_\parallel$ and photon energy $\omega_i$ tuned to the absorption edge $\omega_{\mathrm{edge}}$ and $\omega_0 \sim 0.28$ eV with which the integrated intensity for $\Delta S = 0$ is maximum. RIXS spectra are presented by the calculated spectral weights, i.e., the $\delta$-functions shown by the vertical thin solid lines, after convolution with Lorentzian broadening of $0.2t$. Spectra are vertically offset for clarity. (**c**) Photon energy $\omega_i$ dependence of the integrated RIXS intensity with $\Delta S = 0$ (red) and 1 (blue) at $\mathbf{q}_\parallel = (\frac{\pi}{2}, 0)$, as well as the calculated XAS (gray). $\omega_{\mathrm{edge}}$ and $\omega_0$ are marked by the blue and red arrows, respectively. (**d**) Spin configurations with the largest weights in the XAS edge state indicated by the blue arrow in (**c**). The core holes in the RIXS intermediate state are created in the down-spin sublattice. (**e**) Spin configurations with the largest weights in an eigenstate near the energy indicated by the red arrow in (**c**). The core holes in the RIXS intermediate state are created in the up-spin sublattice.

### Momentum-dependent RIXS

We used RIXS with incident x-ray energies tuned across the Cu $L_3$-edge absorption of energy denoted $L_3$ to study the dispersion of bimagnons. Figure 3a, b show the magnified RIXS spectra and intensity map of $La_2CuO_4$, respectively. In agreement with previous results[19,20,45], phonon and magnon excitations exist for energies smaller than 100 meV and around 280 meV, respectively. The energy of the sharp and well-defined magnon excitation is independent of the incident photon energy; it is a Raman-like excitation. As plotted in Fig. 3a, there is a pronounced spectral feature centered at 0.4 eV on the high-energy side of the magnon excitation for the incident photon energy at $L_3+0.4$ eV, in agreement with the theoretical prediction on the bimagnon RIXS feature[32]. Figure 3c displays the plot of magnon and bimagnon intensities versus the incident photon energy to show their different resonance effects. The bimagnon resonates at an incident photon energy higher than the magnon by 0.3 eV.

The combined results of incident-photon-energy dependent RIXS measurements and ED calculations provide us with an opportunity to study the dynamics of magnon excitation. We set the incident photon energy to 0.6 eV above the $L_3$ absorption edge for measuring the dispersion of bimagnon. Figure 4a shows momentum-dependent RIXS spectra. After the curve fitting analysis shown in the supplementary material, we found that the measured bimagnon disperses from 0.4 eV for $\mathbf{q}_\parallel = (0,0)$ toward 0.5 eV for $\mathbf{q}_\parallel = (\pi,0)$ as plotted in Fig. 4b.

### Discussion

To further understand the bimagnon dynamics of $La_2CuO_4$, we calculated the dynamical bimagnon correlation function, which dominates the non-spin-flip process of the half-filling through its matrix elements[32]. The dynamical bimagnon correlation function is given by a correlation function of the operator $M_\mathbf{q}^\pm = \sum_\mathbf{k}(\cos k_x \pm \cos k_y)\mathbf{S}_{\mathbf{k}+\mathbf{q}} \cdot \mathbf{S}_{-\mathbf{k}}$, where $+$ $(-)$ corresponds to $A_{1g}$ $(B_{1g})$ mode; $\mathbf{S}_{\mathbf{k}+\mathbf{q}}$ and $\mathbf{S}_{-\mathbf{k}}$ are the spins of wave vectors $\mathbf{k}+\mathbf{q}$ and $-\mathbf{k}$, respectively[32]. Figure 5 shows the calculated $\mathbf{q}$-dependent dynamical bimagnon correlation function for $A_{1g}$ and $B_{1g}$ modes. As shown in Fig. 5a, the energy positions of $A_{1g}$ mode indicated by $\delta$-functions have much better correspondence to those in the calculated RIXS spectra with $\Delta S = 0$ in Fig. 2b. In addition, as shown in Fig. 5b, the ED calculations reveal that the bimagnon spectral weight of $B_{1g}$ mode decreases drastically as $\mathbf{q}_\parallel$ is increased toward $(\pi, 0)$, in contrast to our measurement plotted in Fig. 4a. This indicates that the bimagnon of $La_2CuO_4$ detected by Cu $L$-edge RIXS is predominantly of $A_{1g}$ mode.

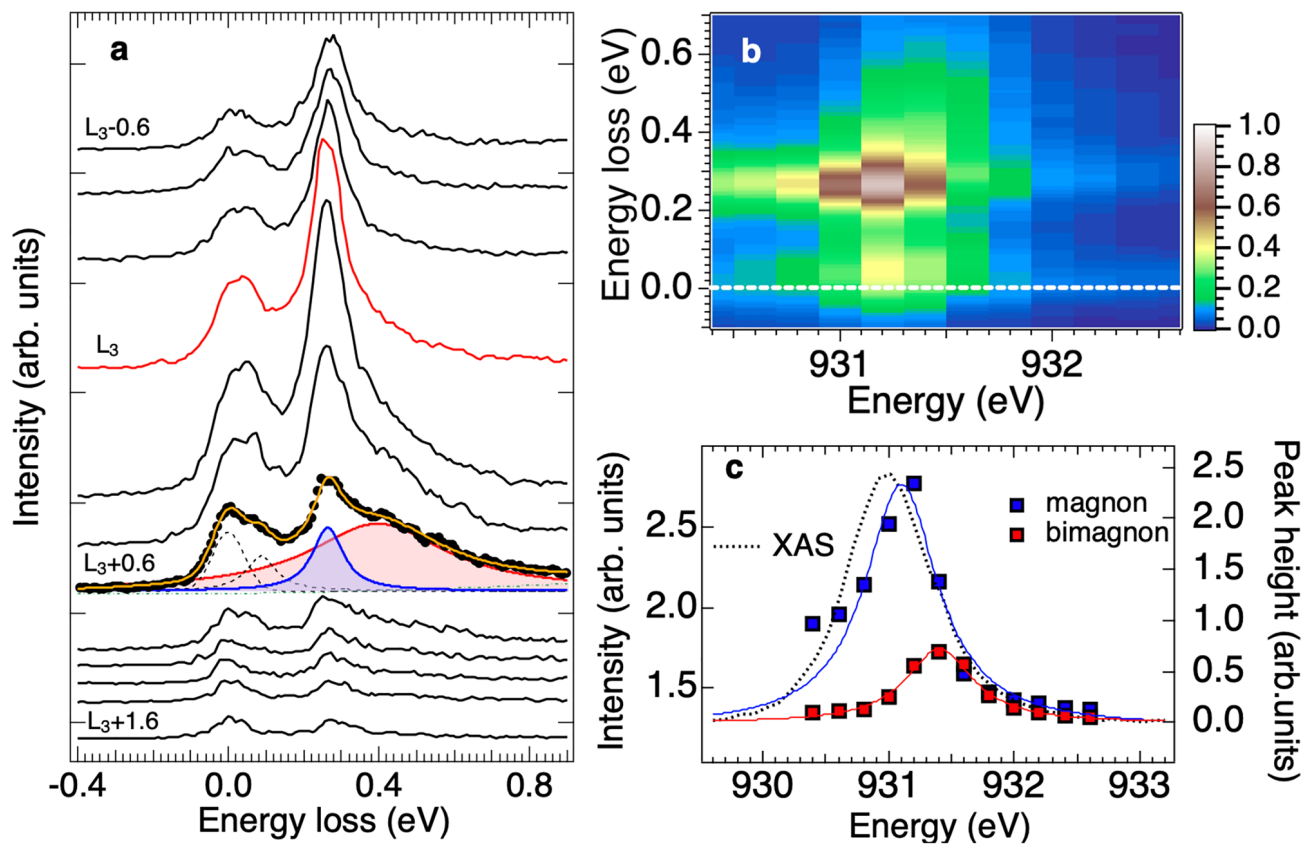

**Fig. 3.** Spin excitation of $La_2CuO_4$ in incident-energy-dependent Cu $L$-edge RIXS. (**a**) Cu $L$-edge RIXS spectra of $\mathbf{q}_\parallel = (0.5\pi, 0)$ for various incident energies across the Cu $L_3$-edge. Spectra are plotted with a vertical offset for clarity. The spectral components of magnon and bimagnon from the curve fitting to the RIXS spectrum induced by x-ray energy of $L_3 + 0.6$ eV are plotted in colors. The dotted lines plot the elastic and the phonon components. See the supplementary material for the curve-fitting details. (**b**) Two-dimensional RIXS intensity map measured across the Cu $L_3$-edge. (**c**) The peak height of the magnon (blue square) and bimagnon (red square) components from the fitting as a function of incident photon energy across the $L_3$-edge. Solid lines plot Lorentizian distribution curves to highlight the resonance along with XAS (gray dashed lines).

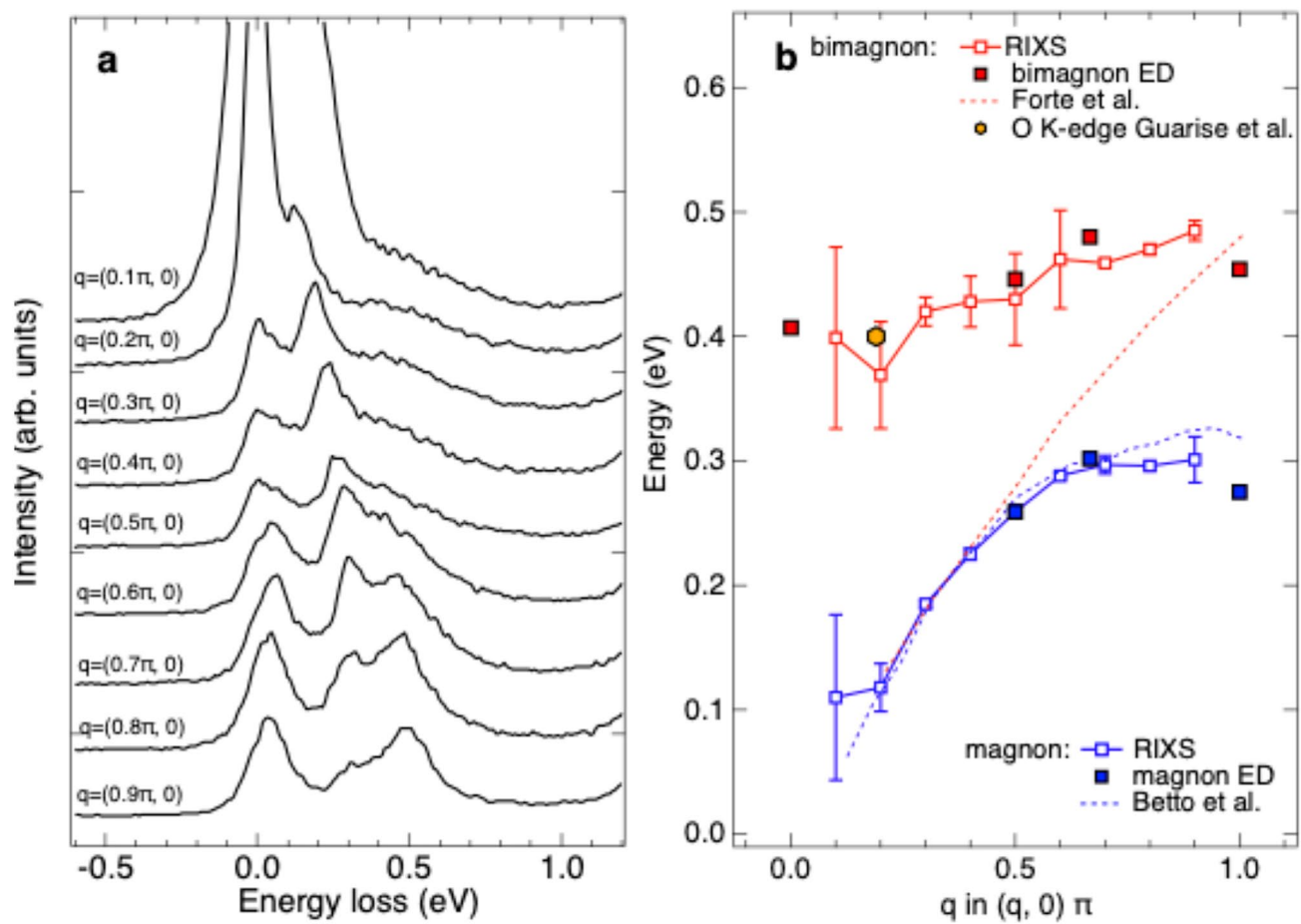

**Fig. 4.** Bimagnon dispersion of $La_2CuO_4$ from momentum-dependent Cu $L$-edge RIXS. (**a**) Cu $L$-edge RIXS spectra for various in-plane momentum transfers with incident energy tuned to $L_3+0.6$ eV. RIXS spectra are after correction for self-absorption and plotted with vertical offsets for clarity. (**b**) Energies of magnon and bimagnon extracted from RIXS, along with the energy positions of the maximum intensities in the calculated RIXS spectra with $\Delta S = 0$ (red squares) and 1 (blue squares). Details of self-absorption correction and curve fitting analysis for determining the energy positions are presented in the supplementary material. For comparison, the energy positions of the single magnon from Betto et al.[46] and the bimagnon from Forte et al.[47] and Guarise et al.[48] are also plotted in (**b**).

The observed bimagnon dispersion presented in Fig. 4b shows a flat-like feature, in sharp contrast to a monotonic and steep dispersion predicted by calculations of the Heisenberg model with only the nearest neighbor exchange interaction[47]. Our observation is consistent with the dispersion in the ED calculations on the Hubbard model. The coherent motion of electrons beyond nearest-neighbor sites and higher-order spin couplings are naturally included in the Hubbard model. This means that the effect of the higher order terms of $t/U$ such as the cyclic interactions is important for the bimagnon dynamics.

The combination of our Cu $L_3$-edge RIXS measurements and ED calculations provides a simple scheme to measure the energy dispersion of bimagnon without resorting to polarization analysis on the scattered x-rays. The incident light was of $\sigma$ polarization, and scattered light of both $\sigma$ and $\pi$ polarization was detected. If the RIXS spectra are recorded at the $L_3$ absorption edge, both channels of $\Delta S = 0$ and $\Delta S = 1$ will be included and the bimagnon excitation will be strongly mixed with single-magnon excitations and the continuum of odd-number magnons[46,49,50]. For the incident x-ray tuned to be above the absorption edge, the bimagnon weight is enhanced. As illustrated in Fig. 2d,e, dominant spin configurations in the RIXS intermediate state having one doubly occupied core-hole site surrounded by spin background are different between RIXS transitions at and above the absorption edge. The latter is more suitable for a coupling to the bimagnon spin configuration in the RIXS final state. This explains why the bimagnon intensity is enhanced. Therefore the measured bimagnon dispersion is more reliable than those extracted from RIXS data recorded with x-rays tuned to the absorption edge.

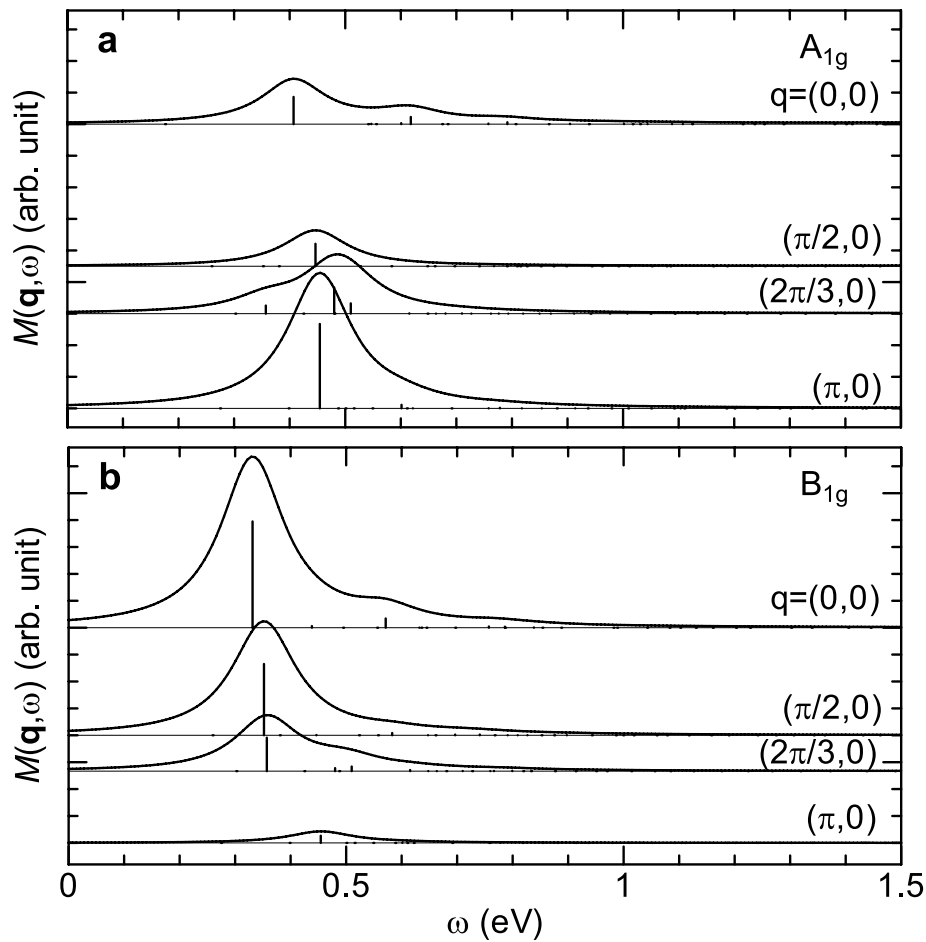

**Fig. 5.** Calculated bimagnon correlations. (a,b) Bimagnon correlations $M(\mathbf{q},\omega)$ with selected q for $A_{1g}$ and $B_{1g}$, respectively. $M(\mathbf{q},\omega)$ were calculated through $\sum_f |\langle f|M_\mathbf{q}^\pm|0\rangle|^2 \delta(\omega - E_f + E_0)$, where $|0\rangle$ and $|f\rangle$ represent, respectively, the ground and final states of energy $E_0$ and $E_f$. The spectra are presented by the calculated spectral weights, i.e., the $\delta$-functions shown by the vertical thin solid lines, after convolution with Lorentzian broadening of $0.2t$. Spectra are vertically offset for clarity.

In short, our RIXS measurements agree well with ED calculations and the results unravel the bimagnon dispersion of $La_2CuO_4$ although the measurement method is straightforward. The effect of the higher-order spin couplings beyond nearest-neighbor sites is important for understanding the bimagnon dynamics of cuprates.

## Methods
### Sample synthesis
The $La_2CuO_4$ single crystal was grown by the traveling-solvent floating zone method[51,52]. After growth, the crystals were annealed appropriately to remove oxygen defects. The oxygen content was tuned to be $4.000\pm0.001$ following Ref.[53]. For more details on the crystal growth and characterization, see Ref.[51].

### RIXS measurements
We conducted RIXS measurements using the AGM-AGS spectrometer of beamline 41A at Taiwan Photon Source[54]. This RIXS beamline has been constructed based on the energy-compensation principle of grating dispersion. The crystallographic axes of the $La_2CuO_4$ crystal were precisely aligned with x-ray Laue diffraction and using a special sample holder with tilting adjustment. Before RIXS measurements, the $La_2CuO_4$ sample was cleaved in air and then mounted on a 3-axis in-vacuum manipulator through a load-lock system. X-ray absorption spectra were measured using a photodiode in the fluorescence yield mode. The sample was cooled to 30 K with liquid helium and RIXS measurements at the Cu $L$-edge were recorded with $\sigma$-polarized incident x-rays for various incident energies with an energy resolution of $\sim 90$ meV FWHM at Cu $L_3$-edge and in-plane wave-vector changes along the anti-nodal direction. The momentum resolution was $\sim 0.006$ Å$^{-1}$ for the $q$-dependent measurements.

### Cu $L$-edge RIXS calculations

We calculated Cu $L$-edge RIXS spectra using the Lanczos-type diagonalization method on a single-band Hubbard model with periodic $4 \times 4$- and $\sqrt{18} \times \sqrt{18}$-site square-lattice clusters. The model Hamiltonian was

$$H_{3d} = -t \sum_{i\delta\sigma} c_{i\sigma}^\dagger c_{(i+\delta)\sigma} - t' \sum_{i\delta'\sigma} c_{i\sigma}^\dagger c_{(i+\delta')\sigma} + U \sum_i n_{i\uparrow} n_{i\downarrow}, \quad (1)$$

where $c_{i\sigma}^\dagger$ is the creation operator of an electron with spin $\sigma$ at site $i$. The number operator $n_{i\sigma}$ is $c_{i\sigma}^\dagger c_{i\sigma}$, and $i+\delta$ ($i+\delta'$) represents the four first (second) nearest-neighbor sites around site $i$. The hopping energy $t$ was a fitting parameter to match the values of the single-magnon energy at $(\frac{\pi}{2}, 0)$. Here the model parameters were set as the second-nearest-neighbor hopping $t'/t = -0.25$, on-site Coulomb interaction $U/t = 10$, and $t = 0.35$ eV. For an energy loss $\Delta\omega$, the RIXS spectra with $\Delta S = 1$ and 0 are given by

$$I_{\mathbf{q}}^{\Delta S=1}(\Delta\omega) = \sum_f |\langle f| S_{\mathbf{q}}^j |0\rangle|^2 \delta(\Delta\omega - E_f + E_0), \quad (2)$$

$$I_{\mathbf{q}}^{\Delta S=0}(\Delta\omega) = \sum_f |\langle f| N_{\mathbf{q}}^j |0\rangle|^2 \delta(\Delta\omega - E_f + E_0), \quad (3)$$

where $|0\rangle$ and $|f\rangle$ represent, respectively, the ground and final states of energy $E_0$ and $E_f$. The operators $S_{\mathbf{q}}^j$ and $N_{\mathbf{q}}^j$ are defined by $S_{\mathbf{q}}^j = (B_{\mathbf{q}\uparrow\uparrow}^j - B_{\mathbf{q}\downarrow\downarrow}^j)/2$ and $N_{\mathbf{q}}^j = (B_{\mathbf{q}\uparrow\uparrow}^j + B_{\mathbf{q}\downarrow\downarrow}^j)$ with

$$B_{\mathbf{q}\sigma'\sigma}^j = \sum_l e^{-i\mathbf{q}\cdot\mathbf{R}_l} c_{l\sigma'} \frac{1}{\omega_i - H_l^j + E_0 + i\Gamma} c_{l\sigma}^\dagger, \quad (4)$$

where $\mathbf{R}_l$ is the position vector at site $l$, $H_l^j = H_{3d} - U_c \sum_\sigma n_{l\sigma} + \varepsilon_j$ with $U_c$ and $\varepsilon_j$ being the Cu $2p$-$3d$ Coulomb interaction and energy level of Cu $2p$ core hole at site $l$, and $j$ is the total angular momentum of the Cu $2p$ hole with $j = 3/2$ for the present $L_3$ edge. The parameter for the core-hole lifetime was set as $\Gamma/t = 1$, $U_c/t = 12$, and the value of $\varepsilon_j$ was determined to match the energy of the edge in the XAS. The polarization dependence of the incident and scattered photons was not included in the calculation of RIXS spectral weight. See Ref.[32] for other calculation details. Correspondingly, the Cu $L$-edge XAS spectrum is also defined in Ref.[32], where $\Gamma_{\text{XAS}} = \Gamma$.

### Data availability

All data generated or analysed during this study are included in this published article [and its supplementary information files].

### References

1. Scalapino, D. J., Loh, E. & Hirsch, J. E. $d$-wave pairing near a spin-density-wave instability. *Phys. Rev. B* **34**, 8190–8192 (1986).
2. Mook, H. A., Yethiraj, M., Aeppli, G., Mason, T. E. & Armstrong, T. Polarized neutron determination of the magnetic excitations in $YBa_2Cu_3O_7$. *Phys. Rev. Lett.* **70**, 3490–3493 (1993).
3. Fong, H. et al. Neutron scattering from magnetic excitations in $Bi_2Sr_2CaCu_2O_{8+\delta}$. *Nature* **398**, 588–591 (1999).
4. Tsuei, C. C. & Kirtley, J. R. Pairing symmetry in cuprate superconductors. *Rev. Mod. Phys.* **72**, 969–1016 (2000).
5. Maier, T. A., Jarrell, M. S. & Scalapino, D. J. Structure of the pairing interaction in the two-dimensional Hubbard model. *Phys. Rev. Lett.* **96**, 047005 (2006).
6. Scalapino, D. J. A common thread: The pairing interaction for unconventional superconductors. *Rev. Mod. Phys.* **84**, 1383 (2012).
7. Fradkin, E., Kivelson, S. A. & Tranquada, J. M. Colloquium: Theory of intertwined orders in high temperature superconductors. *Rev. Mod. Phys.* **87**, 457–482 (2015).
8. Lorenzana, J. & Sawatzky, G. A. Phonon assisted multimagnon optical absorption and long lived two-magnon states in undoped lamellar copper oxides. *Phys. Rev. Lett.* **74**, 1867 (1995).
9. Lorenzana, J. & Sawatzky, G. A. Theory of phonon-assisted multimagnon optical absorption and bimagnon states in quantum antiferromagnets. *Phys. Rev. B* **52**, 9576 (1995).
10. Lorenzana, J., Eroles, J. & Sorella, S. Does the Heisenberg model describe the multimagnon spin dynamics in antiferromagnetic CuO layers?. *Phys. Rev. Lett.* **83**, 5122 (1999).
11. Manousakis, E. The spin-1/2 Heisenberg antiferromagnet on a square lattice and its application to the cuprous oxides. *Rev. Mod. Phys.* **63**, 1–62 (1991).
12. Coldea, R. et al. Spin waves and electronic interactions in $La_2CuO_4$. *Phys. Rev. Lett.* **86**, 5377 (2001).
13. Headings, N. S., Hayden, S. M., Coldea, R. & Perring, T. G. Anomalous high-energy spin excitations in the high-$T_c$ superconductor-parent antiferromagnet $La_2CuO_4$. *Phys. Rev. Lett.* **105**, 247001 (2010).
14. Manojlović, M. et al. Spin-wave dispersion and transition temperature in the cuprate antiferromagnet $La_2CuO_4$. *Phys. Rev. B* **68**, 014435 (2003).
15. MacDonald, A. H., Girvin, S. M. & Yoshioka, D. $\frac{t}{U}$ expansion for the Hubbard model. *Phys. Rev. B* **37**, 9753–9756 (1988).
16. Peres, N. M. R. & Araújo, M. A. N. Spin-wave dispersion in $La_2CuO_4$. *Phys. Rev. B* **65**, 132404 (2002).
17. Dalla Piazza, B. et al. Unified one-band Hubbard model for magnetic and electronic spectra of the parent compounds of cuprate superconductors. *Phy. Rev. B* **85**, 100508 (2012).
18. Kastner, M. A., Birgeneau, R. J., Shirane, G. & Endoh, Y. Magnetic, transport, and optical properties of monolayer copper oxides. *Rev. Mod. Phys.* **70**, 897–928 (1998).

19. Braicovich, L. et al. Dispersion of magnetic excitations in the cuprate $La_2CuO_4$ and $CaCuO_2$ compounds measured using resonant x-ray scattering. *Phys. Rev. Lett.* **102**, 167401 (2009).
20. Braicovich, L. et al. Magnetic excitations and phase separation in the underdoped $La_{2-x}Sr_xCuO_4$ superconductor measured by resonant inelastic x-ray scattering. *Phys. Rev. Lett.* **104**, 077002 (2010).
21. Braicovich, L. et al. Momentum and polarization dependence of single-magnon spectral weight for Cu $L_3$-edge resonant inelastic x-ray scattering from layered cuprates. *Phys. Rev. B* **81**, 174533 (2010).
22. Ament, L. J. P., van Veenendaal, M., Devereaux, T. P., Hill, J. P. & van den Brink, J. Resonant inelastic X-ray scattering studies of elementary excitations. *Rev. Mod. Phys.* **83**, 705 (2011).
23. Dean, M. P. M. et al. Spin excitations in a single $La_2CuO_4$ layer. *Nat. Mater* **11**, 850–854 (2012).
24. Ivashko, O. et al. Strain-engineering Mott-insulating $La_2CuO_4$. *Nat. Commun.* **10**, 1–8 (2019).
25. Robarts, H. C. et al. Dynamical spin susceptibility in $La_2CuO_4$ studied by resonant inelastic x-ray scattering. *Phys. Rev. B* **103**, 224427 (2021).
26. de Groot, F. M. F., Kuiper, P. & Sawatzky, G. A. Local spin-flip spectral distribution obtained by resonant X-ray Raman scattering. *Phys. Rev. B* **57**, 14584–14587 (1998).
27. Ament, L. J. P., Ghiringhelli, G., Sala, M. M., Braicovich, L. & van den Brink, J. Theoretical demonstration of how the dispersion of magnetic excitations in cuprate compounds can be determined using resonant inelastic x-ray scattering. *Phys. Rev. Lett.* **103**, 117003 (2009).
28. Haverkort, M. W. Theory of resonant inelastic X-ray scattering by collective magnetic excitations. *Phys. Rev. Lett.* **105**, 167404 (2010).
29. Igarashi, J.-I. & Nagao, T. Magnetic excitations in $L$-edge resonant inelastic X-ray scattering from cuprate compounds. *Phys. Rev. B* **85**, 064421 (2012).
30. Jia, C. J. et al. Persistent spin excitations in doped antiferromagnets revealed by resonant inelastic light scattering. *Nat. Commun.* **5**, 1–7 (2014).
31. Jia, C., Wohlfeld, K., Wang, Y., Moritz, B. & Devereaux, T. P. Using RIXS to uncover elementary charge and spin excitations. *Phys. Rev. X* **6**, 021020 (2016).
32. Tsutsui, K. & Tohyama, T. Incident-energy-dependent spectral weight of resonant inelastic X-ray scattering in doped cuprates. *Phys. Rev. B* **94**, 085144 (2016).
33. Huang, H. Y. et al. Raman and fluorescence characteristics of resonant inelastic X-ray scattering from doped superconducting cuprates. *Sci. Rep.* **6**, 1–7 (2016).
34. Chaix, L. et al. Dispersive charge density wave excitations in $Bi_2Sr_2CaCu_2O_{8+\delta}$. *Nat. Phys.* **13**, 952–956 (2017).
35. Huang, H. Y. et al. Jahn-Teller distortion driven magnetic polarons in magnetite. *Nat. Commun.* **8**, 15929 (2017).
36. Hepting, M. et al. Three-dimensional collective charge excitations in electron-doped copper oxide superconductors. *Nature* **563**, 374–378 (2018).
37. Huang, H. Y. et al. Quantum fluctuations of charge order induce phonon softening in a superconducting cuprate. *Phys. Rev. X* **11**, 041038 (2021).
38. Huang, H. Y. et al. Resonant inelastic X-ray scattering as a probe of $j_{eff}=1/2$ state in 3 D transition-metal oxide. *npj Quantum Mater.* **7**, 33 (2022).
39. Singh, A. et al. Acoustic plasmons and conducting carriers in hole-doped cuprate superconductors. *Phys. Rev. B* **105**, 235105 (2022).
40. Singh, A. et al. Unconventional exciton evolution from the pseudogap to superconducting phases in cuprates. *Nat. Commun.* **13**, 7906 (2022).
41. Li, Q. et al. Prevailing charge order in overdoped La 2−x Sr x CuO 4 beyond the superconducting dome. *Phys. Rev. Lett.* **131**, 116002 (2023).
42. Bisogni, V. et al. Bimagnon studies in cuprates with resonant inelastic x-ray scattering at the O $K$-edge. I. Assessment on $La_2CuO_4$ and comparison with the excitation at Cu $L_3$ and Cu $K$ edges. *Phys. Rev. B* **85**, 214527 (2012).
43. Devereaux, T. P. et al. Directly characterizing the relative strength and momentum dependence of electron-phonon coupling using resonant inelastic X-ray scattering. *Phys. Rev. X* **6**, 041019 (2016).
44. Lin, J. Q. et al. Strongly correlated charge density wave in $La_{2-x}Sr_xCuO_4$ evidenced by doping-dependent phonon anomaly. *Phys. Rev. Lett.* **124**, 207005 (2020).
45. Chaix, L. et al. Resonant inelastic x-ray scattering studies of magnons and bimagnons in the lightly doped cuprate $La_{2-x}Sr_xCuO_4$. *Phys. Rev. B* **97**, 155144 (2018).
46. Betto, D. et al. Multiple-magnon excitations shape the spin spectrum of cuprate parent compounds. *Phys. Rev. B* **103**, L140409 (2021).
47. Forte, F., Ament, L. J. & van den Brink, J. Magnetic excitations in $La_2CuO_4$ probed by indirect resonant inelastic x-ray scattering. *Phys. Rev. B* **77**, 134428 (2008).
48. Guarise, M. et al. Measurement of magnetic excitations in the two-dimensional antiferromagnetic Sr 2 CuO 2 Cl 2 insulator using resonant X-ray scattering: Evidence for extended interactions. *Phys. Rev. Lett.* **105**, 157006 (2010).
49. Sandvik, A. W. & Singh, R. R. P. High-energy magnon dispersion and multimagnon continuum in the two-dimensional Heisenberg antiferromagnet. *Phys. Rev. Lett.* **86**, 528 (2001).
50. Martinelli, L. et al. Fractional spin excitations in the infinite-layer cuprate $CaCuO_2$. *Phys. Rev. X* **12**, 021041 (2022).
51. Komiya, S., Ando, Y., Sun, X. F. & Lavrov, A. N. $c$-axis transport and resistivity anisotropy of lightly to moderately doped $La_{2-x}Sr_xCuO_4$ single crystals: Implications on the charge transport mechanism. *Phys. Rev. B* **65**, 214535 (2002).
52. Komiya, S., Chen, H.-D., Zhang, S.-C. & Ando, Y. Magic doping fractions for high-temperature superconductors. *Phys. Rev. Lett.* **94**, 207004 (2005).
53. Kanai, H. et al. Defect chemistry of $La_{2-x}Sr_xCuO_{4-\delta}$: Oxygen nonstoichiometry and thermodynamic stability. *J. Solid State Chem.* **131**, 150–159 (1997).
54. Singh, A. et al. Development of the soft X-ray AGM-AGS RIXS beamline at the Taiwan photon source. *J. Synchrotron Radiat.* **28**, 977 (2021).

### Acknowledgements
This work was partly supported by the NSCT of Taiwan under Grant Nos. 109-2112-M-213-010-MY3, 109-2923-M-213-001, and 113-2112-M-007-033 by JSPS under Grant Nos. JP19K03741, JP22K03535, JP22K03500, JP19H01829, JP19H05825, 24K00560, by MEXT under the "Program for Promoting Researches on the Supercomputer Fugaku" (Basic Science for Emergence and Functionality in Quantum Matter, No. JPMXP1020200104) from MEXT, and by QST President's Strategic Grant (QST Advanced Study Laboratory). Part of the computational work was performed using the supercomputing facilities in QST. A.S. was partially supported by ANRF (SERB) under grant number 2023/000242. A.F. acknowledges the support from the Yushan Fellow Program and the Center for Quantum Science and Technology within the framework of the Higher Education Sprout Project under the Ministry of Education of Taiwan.

### Author contributions

A.S., H.Y.H., J.O., D.J.H., and C.T.C. developed the RIXS instruments and conducted the RIXS experiments. A.S., H.Y.H., and D.J.H. analyzed the data. K.T. and T.T. performed RIXS calculations. S.K. synthesized and characterized the sample. D.J.H., A.S.,A.F., K.T., and T.T. wrote the paper with other authors' input.

### Declarations

### Competing interests

The authors declare no competing interests.

### Additional information

**Supplementary Information** The online version contains supplementary material available at https://doi.org/10.1038/s41598-025-15435-5.

**Correspondence** and requests for materials should be addressed to D.J.H.

**Reprints and permissions information** is available at www.nature.com/reprints.

**Publisher's note** Springer Nature remains neutral with regard to jurisdictional claims in published maps and institutional affiliations.